# Oxygen Octahedral Tilt Controlled Topological Hall Effect in Epitaxial and Freestanding SrRuO$_3$/SrIrO$_3$ Heterostructures

Z. S. Lim, A. K. H. Khoo, Z. Zhou, G. J. Omar, P. Yang, R. Laskowski, A. Ariando*

Dr. Z. S. Lim, Z. Zhou, Dr. G. J. Omar, Prof. A. Ariando
Department of Physics, 2 Science Drive 3, National University of Singapore 117551.
#11-01, NUSNNI, T-Lab, 5A, Engineering Drive 1, National University of Singapore 117411.
email: phyarian@nus.edu.sg

Dr. A. K. H. Khoo, Dr. R. Laskowski
Institute of High-Performance Computing, A*STAR, 1 Fusionopolis Way, #16-16 Connexis, Singapore 138632

Dr. P. Yang
Singapore Synchrotron Light Source, 5 Research Link, NUS 117603.

**Abstract:**

**The fabrication technique of oxide flakes by epitaxial lift-off has allowed the investigation of a freestanding version of multifunctional oxide thin films and heterostructures. Several recent reports have demonstrated the robustness of these freestanding oxides retaining their desirable properties after detachment from the substrate. Here, we first demonstrate various epitaxial SrRuO$_3$/SrIrO$_3$ heterostructures showing Hall-humps reminiscence of Topological Hall Effect. First-principle calculations reveal that octahedral tilts can modify the sign of the Dzyaloshinskii-Moriya interaction and avoid cancellation in a symmetric trilayer structure. Secondly, freestanding flakes of the trilayer can also retain the Hall-humps across a wide temperature range. The behavior of humps' peak field does not vary with field direction rotation away from the surface normal, consistent with a micromagnetic simulation result of Néel-type magnetic Skyrmions. The layer-resolved octahedral tilts in the freestanding heterostructures also crucially control the occurrence of Hall-humps, consistent with the insight from the epitaxial ones. This work offers a new perspective to understanding Hall-humps in perovskite oxides, as well as demonstrates the fabrication of oxide heterostructure membranes with high interfacial quality.**

**Main Text:**

Oxide materials encompass various functional classes such as ferroelectric[1], magnetic superexchange[2], photovoltaics[3], multiferroics with magnetoelectric coupling[4], high-temperature superconductivity[5], strongly correlated high-mobility two-dimensional electron liquid (2DEL)[6], pyrochlore Weyl semimetal[7], topology-related magnetic[8] and polar[9] textures. From the aspect of practical device fabrications, besides being more stable in air than other exotic materials, perovskite oxides can achieve high crystallinity and sharp interfaces by thin-film molecular epitaxy. However, integrating high quality oxides with the inexpensive silicon platform in industry presents formidable challenges, due to mismatch in crystal structures and unwanted formation of $SiO_X$[10]. One interesting route to the mentioned integration is by epitaxial lift-off (ELO), after a lattice-matched sacrificial layer such as $Sr_3Al_2O_6$ (SAO)[11] or $La_{0.67}Sr_{0.33}MnO_3$ (LSMO)[12] is dissolved in water or etchant, respectively. The freestanding flakes can be transferred onto other inexpensive supports, while the original lattice-matched substrates can even be reused after film detachment. Such a technique could enable exploration into Van der Waals forces (VdW) related new physics for oxide thin films similar to that of 2D materials [13]. Alternatively, probe-beam transmission mode magnetic imaging[14] and ultrafast pump-probe optical excitation of Skyrmions breathing/rotational modes[15] could also be done on freestanding thin films. Recently, there have been numerous reports demonstrating fabrication of freestanding oxides with their desirable properties well retained or even improvised, such as few-unit-cell ferroelectricity[16], tunable photovoltaic by bending strain via flexoelectricity[17], or enhanced saturation magnetization[11].

On the other hand, the ultrathin $SrRuO_3/SrIrO_3$ (SRO/SIO) heterostructures were recently discovered to show interesting Topological Hall Effect (THE)[18] with rich evolutions, thanks to the presence of strong spin-orbit coupling (SOC) and sharp interface. THE has been theorized to emerge when free electrons travel across non-collinear magnetic textures with a net topological charge $Q = \frac{1}{4\pi} \int \widehat{m} \cdot \left( \frac{d\widehat{m}}{dx} \times \frac{d\widehat{m}}{dy} \right) dxdy$ and receive a perpendicular deflection[19]. Unfortunately, the hump-shape Hall features observed in several material systems could be associated with either true magnetic Skyrmions[20], double-*q* system of incommensurate spin-crystal[21] or inhomogeneous domains with opposite-sign *k*-space Berry curvatures[22, 23] – the ambiguity remains unclear and non-universal among different cases. Notably, that real Skyrmions satisfy the criterion of negative domain wall energy density $\sigma_{DW} = \sqrt{4A_{ex}K_u} - D < 0$ where $A_{ex}$, $K_u$ and $D$ are the exchange stiffness, uniaxial anisotropy and Dzyaloshinskii-Moriya Interaction (DMI), respectively[24], yet regular collinear magnetic domains may not. Henceforth, the THE-like signals are referred as "Hall-humps". In this work, we studied the effects of ELO with SAO buffer on the Hall-humps of several SRO/SIO combinations. Such structures and their Hall Effects are known to be extremely sensitive to various perturbations, hence fabricating freestanding flakes while retaining the Hall-humps signal should be a challenging task. Then, we employed a magnetic field rotation technique to show that the Hall-humps' behaviour agrees with magnetic Skyrmions' topological charge density from a micromagnetic simulation. Besides, we found a crucial linkage between the existence of the Hall-humps and $a^-/b^-$ type oxygen octahedral tilts, which should be a novel finding.

First, we compared the properties among several epitaxial structures. Structure A with SIO(10uc) on top of SRO(5uc) has been similarly investigated in past publications, while structure B has the inverted layer stacking sequence as A. Both structures were grown assisted by monitoring the Reflection High Energy Electron Diffraction (RHEED), with data provided in supporting **Figure S1**a. In **Figure 1**, the Anomalous Hall Effect (AHE) and Hall-humps are present in structures A and B, yet their low temperature (ground state) AHE has opposite signs. The low-temperature negative AHE in structure A is consistent with the reported behaviors for SIO/SRO bilayers and SRO single layer[18, 25]. This can likely be attributed to the bulk-like ferromagnetic minority-spin double-exchange (FmDE) mechanism with threefold orbital-degeneracy in SRO (Figure 1c, left)[26]. In FmDE, the mobile electron's spin antiparallel ($t_{2g}^\downarrow$) to the local magnetization produces a negative spin polarization $P_S = \frac{DOS_\uparrow(E_F) - DOS_\downarrow(E_F)}{DOS_\uparrow(E_F) + DOS_\downarrow(E_F)}$ (-9.5% at 0.31 K)[27] at Fermi level ($E_F$) and Weyl nodes with negative Chern numbers ($C_n$)[28]. However, supporting **Figure S1**b shows that structure A (with direct SRO/STO interface) has lower Curie temperature ($T_C$) and saturation magnetization ($M_{sat}$) than structure B (with SIO inserted between SRO and STO). This suggests a significant portion of bottommost SRO in structure A has weakened FmDE due to the suppression of octahedral tilts by the cubic substrate[29]. We may also expect the confinement effect in (001) to lift the $t_{2g}$ orbitals' degeneracy by lowering the $d_{xy}$ orbital energy than $d_{xz,yz}$ orbitals due to heavier z-direction electron effective mass ($m_z^*$)[30]. This promotes the antiferromagnetic superexchange (AFSE) (Figure 1c, middle) in competition with the FmDE, hence reduces its $T_C$ and $M_{sat}$. By buffering the cubic substrate with SIO(10uc) in structure B, a good SRO/SIO interface can be expected. Additional contributions of $Ir^{4+}$ to $Ru^{4+}$ electron transfer, ferrimagnetic $Ir^{(4+\delta)+}$–$O^{2-}$–$Ru^{(4-\delta)+}$ superexchange (Figure 1c, right) and electronic bands with dominantly positive $C_n$ have also been predicted in a few first-principle calculations[22, 31]. In structure C, we see a near-complete cancelling of AHE but strong Hall-humps up to 100 K, which can be understood as a summation of $\rho_{xy}$ from structures A and B.

However, it is ambiguous whether THE humps present mutual enhancement or cancellation in structure C. In metallic systems, two reciprocal interfaces are expected to have opposite-sign DMI vectors[32]. Recently in centrosymmetric oxides, strain gradient due to relaxation across large thicknesses was identified theoretically and experimentally as the source of DMI for stabilizing magnetic Skymions[33]. To gain deeper insights, we performed Density Functional Theory (DFT) calculations on structures A and B by accounting for details of experimentally measured octahedral tilts. **Figure 2**a shows the half-integer HKL X-ray Bragg diffraction data at (0, 0.5, 1.5), (0.5, 0, 1.5), (0.5, 1.5, 1), (0.5, 0.5, 1.5), (1.5, 0.5, 1.5) and (0.5, 1.5, 1.5) as evidences for $a^+$, $b^+$, $c^+$, $a^-/b^-$, $a^-/c^-$ and $b^-/c^-$ types of tilts respectively, according to Glazer's notations[34]. Bulk orthorhombic crystals with space group Pbnm have Glazer notation of $a^-a^-c^+$, but their epitaxial films on in-plane compressive STO(001) substrate would produce $a^-b^+c^-$ or $a^+b^-c^-$ domains. This is because the B-O-B bonds along the out-of-phase (-) tilt axes are longer than that along in-phase (+) axes; hence it is not energetically favorable to have out-of-phase tilts along both the compressed in-plane directions[35]. Such rule-of-thumb applies to relatively thick SIO and SRO films since their lattice parameters are close. Besides, the octahedral tilts of both materials will be restricted in their proximity regime[29] by the cubic STO substrate. Hence for structure B, the tilting pattern of the bottom SIO(10uc) gradually evolves from $a^0a^0c^0$ to $a^+b^0c^0$ after exceeding a critical thickness, followed by $a^-b^+c^0$ for the top SRO (Figure 2c). However, the bottom SRO(5uc) in structure A does not exceed the critical thickness, and the octahedra are best described as tetragonal (restricted) around the interface (Figure 2b).

Using a two-layer perovskite model bordered by vacuum for DFT, i.e., $a^0b^0c^0$(SIO)/$a^0b^0c^0$(SRO) for structure A and $a^-b^+c^0$(SRO)/$a^+b^0c^0$(SIO) for structure B, we investigated the atomic-layer resolved DMI around the interface[36], as described in *Methods*. Since cycloidal magnons but not Skyrmions can exist as ground state suitable for DFT calculations, the cycloidal wavevector $q = \frac{2\pi}{L} = \frac{\beta}{a}$ was defined on $Ru^{4+}$ along the a-direction and gradually varied to find the total energy minimum, where *L*, *β*, and *a* are the spin wavelength, canting angle of neighboring $Ru^{4+}$ and in-plane pseudocubic lattice parameter respectively. Assuming that the role of uniaxial anisotropy is negligible at ultrathin regime, the total energy, $E_{tot} = C_{ex}q^2 + D_{ind}q$ is the summation of $q^2$-dependent exchange stiffness and

linear-$q$ DMI energies, while the fitting constant $D_{ind}$ is equivalent to the effective DMI vector magnitude in the atomistic definition, $E_{DMI} = \sum_{ij} \boldsymbol{D} \cdot (\hat{\boldsymbol{s}}_i \times \hat{\boldsymbol{s}}_j)$. In Figure 2d-e, the layer-resolved DMI contributed by Ir- and Ru-atoms shows sign-reversal between structures A and B and magnitude adjustments, while the O-atoms in the $RuO_2$-plane gain additional DMI and those in the $IrO_2$-plane maintained the same sign. This way, in structures A and B, the total $D_{ind}$ values are -2.43 mJ/m$^2$ and +0.76 mJ/m$^2$, respectively, resulting in the optimized $q$=0.128 Å$^{-1}$ and -0.031 Å$^{-1}$ (opposite wavevectors) corresponding to β=28.8° and -6.9° respectively after total energy minimization. Straightforwardly, structure C can be estimated to have $D_{ind}$~-1.67 mJ/m$^2$ suitable for Skyrmion stabilization at small magnetic fields. This group of epitaxial structures A, B and C form the basis of our study, to be compared with their freestanding counterparts A$_f$, B$_f$, and C$_f$ (**Figure 3**).

Next, the quality of the freestanding structure was carefully optimized. Preliminary experiments indicated prolonged exposure to water might deteriorate the SRO/SIO sample quality, probably due to oxygen vacancy migration. Hence, a good strategy is to minimize the float time of samples in water (5-10 minutes) before scooping out onto respective supports. This requires the SAO sacrificial layer to be thick enough (100 nm) for fast water dissolution, maintaining layer-by-layer growth with a surface roughness of ~0.16 nm, which mandates a high base-vacuum of <5x10$^{-6}$ Torr during the SAO film growth. However, due to lattice mismatch with SAO, a buffer layer of 40nm-thick STO should be inserted after SAO to avoid island growth of SRO and SIO. Its thickness was optimized to produce elongated flakes convenient for photolithography and Hall bar fabrication, yet avoid rolling due to imbalanced surface tension (**Figure S2**a,b)[37]. Finally, after transferring onto the $SiO_2$//Si substrates (Figure 3b), the flakes maintained smooth topography of ~0.23 nm roughness.

To obtain a sufficiently strong signal in magnetometry measurement, we exfoliated a large (mm-scale) flake of structure C$_f$ by attaching the sample surface to a dicing tape before immersing into water; nevertheless, the large flake contains abundant cracks due to strain relaxation (Figure S2c). In Figure S2d, a clear $T_C$~100 K can be inferred from the moment versus temperature (*M-T*) curves, where the bifurcation between field-cooled (FC) and zero-field-cooled (ZFC) *M-T* curves is similar to SRO films having a competition between ferromagnetic and antiferromagnetic interaction between Ru$^{4+}$. In the moment-versus-field (*M-H*) loops, clear double-hysteresis loops can be seen at a low temperature of 20 K, with the wider coercive field μ$_o H_C$~1 T, which is similar to the $H_C$ of Hall Effect in Figure 3c. Thus the wider $H_C$ can be attributed to the well-known Ru$^{4+}$/Ru$^{4+}$ ferromagnetic exchange, while the narrower $\mu_oH_C$~28 mT originates from $Ru^{4+}/Ir^{4+}$ superexchange at the interface, weakly-magnetized $Ir^{4+}$ moment by proximity effect with SRO, or oxygen vacancy in the 40-nm STO buffer. The bigger $H_C$ quickly diminishes with temperature and becomes indistinguishable at ~60 K, but the thinner hysteresis loop is almost temperature-insensitive and persists up to 300 K. Clear out-of-plane magnetic anisotropy can be seen in both *M-T* and *M-H* curves. In comparison to SAO, the $La_{0.67}Sr_{0.33}MnO_3$ (LSMO), which is also a candidate as a sacrificial layer to be etched by a "KI+HCl+$H_2O$" reducing agent[12, 38], is not suitable in our case due to the fast chemical reaction of SRO and SIO films with the acidic etchant and become damaged (Figure S2e,f).

To measure electrical properties, Cu electrodes were patterned onto freestanding flakes $A_f$, $B_f$ and $C_f$ by standard photolithography and lift-off (Figure 3b). From Figure 3c, structure $A_f$ ($B_f$) is similar to A (B) in terms of their respective negative (positive) AHE signs. Yet, the Hall-humps are suppressed at all temperatures, and structure $B_f$ has slightly larger $H_C$ than that of structures B, A and $A_f$ at 10 K, indicating the inevitable deterioration of the interface quality. Surprisingly, the Hall-humps can be observed in structure $C_f$ in the range of 10-80 K, suggesting DMI survival is achievable, together with the dominant negative-sign AHE component. It is vital to keep the water immersion time short; failure to do so would lead to suppression of THE signal at higher temperature (Figure 3d), to be correlated with a weakened interface DMI.

Without high-resolution magnetic imaging, drawing relationships between Hall-humps and magnetic Skyrmions has been difficult. Some earlier publications attributed the non-reversible behavior in minor loop measurement up to the Hall-humps' peak field ($H_{hump}$) as counter-evidence for THE of magnetic Skyrmions[39]. Considering that THE can be hysteretic[18, 40] or non-hysteretic[41] depending on how close the Skyrmions occur near-zero magnetic field, the claim mentioned is also debatable and inconclusive. We illustrate this argument in **Figure S4**a. Here we offer a more convincing method to investigate the Hall-humps' behavior by performing rotation of magnetic field from out-of-plane ($H\|z$, $\theta=0º$) to near in-plane ($H\|\sim y$, $\theta=80º$) on structure $C_f$ at 80 K. As shown in Figure 3e,f, the Hall-humps exist up to $\theta=70º$, yet the peak $H_{hump}$ remains almost unchanged; while the apparent $H_C$ increases following the $1/\cos(\theta)$ trend well up to $\theta=60º$. This implies that, in contrast to $H_C$ that is sensitive to the z-component of the magnetic field, $H_{hump}$ is sensitive to the total field magnitude. We speculate that $H_C$ and $H_{hump}$ originate from the bottom and top SRO layers, respectively. For comparison, we further performed a $MUMAX^3$ micromagnetic simulation[42] on true Néel-type Skyrmions (Figure 3g), with more information given in Supplementary text and **Figure S3**. Reasonable values of $A_{ex}$, $K_u$ and $M_{sat}$ extracted from previous publications were used, and the interface

DMI ($D_{ind}$) was adjusted such that the topological charge density, $TCD = \boldsymbol{m} \cdot \left(\frac{\partial \boldsymbol{m}}{\partial x} \times \frac{\partial \boldsymbol{m}}{\partial y}\right)$ reaches maximum at $\mu_o H = \sim 0.15$ T, matching with $H_{hump}$ in Figure 3e,f at $\theta=0$. Fascinatingly, the peak $TCD$ shown in the $\theta$-$\mu_o H$ mapping stays at roughly constant $\mu_o H$ before completely vanishing at large $\theta$, agreeing well with the measured constant $H_{hump}$ behavior, where $\mu_o H$ is the total magnetic field. Such behavior can be understood from the Ginzburg-Landau framework – a hexagonal Skyrmion-lattice is formed by a trio of magnetic helicoids/cycloids propagating in 120º from each other in-plane (3q-lattice)[43], namely: $\boldsymbol{m}_i = \sum_i^{1,2,3}[\boldsymbol{m}_z \cos(\boldsymbol{q}_i \cdot \hat{\boldsymbol{r}}) \pm (\boldsymbol{m}_z \times \boldsymbol{q}_i) \sin(\boldsymbol{q}_i \cdot \hat{\boldsymbol{r}})]$ for Bloch-type and $\boldsymbol{m}_i = \sum_i^{1,2,3}[\boldsymbol{m}_z \cos(\boldsymbol{q}_i \cdot \hat{\boldsymbol{r}}) \pm \sin(\boldsymbol{q}_i \cdot \hat{\boldsymbol{r}})]$ for Néel-type, where $\boldsymbol{m}_z$=[001]. For thin films, if $H_z^*$ alone is an out-of-plane field suitable to produce the maximum Skyrmion density, an in-plane $H_y$ added to $H_z^*$ would destroy the Skyrmion-lattice but elongate them into helical/cycloidal stripes[44] (Figure S4c), hence the total field $H_T = \sqrt{H_z^2 + H_y^2}$ to achieve the maximum Skyrmion density should not grow with $1/\cos(\theta)$ which is the case where $H_y$ plays no role. For Bloch-type Skyrmions in bulk chiral materials, tilting magnetic field will not destroy the Skyrmions but their core would align with $H_T$, with $H_{maxTCD}$ remains fixed. These arguments support that structure $C_f$ hosts Néel-type Skyrmions.

Figure 4a shows the layer-resolved half-integer HKL XRD data of the STO/SAO buffered heterostructures from the first to the fifth layers to establish a consistent connection to the first half of this paper. Due to difficulty in preparing continuous large-area, high-quality freestanding membrane for precise crystallographic alignment, these data were obtained without water dissolution of the SAO, but care was taken to perform the XRD measurement immediately after sample growth to prevent deterioration by air moisture. Since SAO's lattice parameters are approximately four times that of STO, the strong (026) and (206) peaks of SAO inevitably appear at the search for $a^+$ and $b^+$. The STO/SAO (blue curves) and the SRO/STO/SAO (green curves) indicate that the STO buffer and the bottom SRO contribute no tilts ($a^0a^0c^0$). Continuing with SIO/SRO/STO/SAO (orange curves), $a^+$ and $b^+$ tilts do not emerge. Finally, the $a^-/b^-$ tilt emerges (red curves) in the top SRO when all five layers are completed (structure $C_f$), accompanied by the emergence of the Hall-humps. Conversely, if the bottom SRO is removed to form structure $B_f$ – SRO/SIO/STO/SAO (yellow curves), the top SRO can still exhibit the $a^-/b^-$ tilt as distinct to structure B, resulting in the absence of the Hall-hump in structure $B_f$. This suggests the suppression of octahedral tilt to a more severe extent by the grown STO buffer compared to the substrate or presence of high defect density. In summary, such insight leads to the structural sketches of the freestanding SRO/SIO/SRO flakes in Figure 4b-e, showing $a^-b^0c^0$ for the top SRO, $a^+b^+c^0$ for SIO near the top interface, and $a^0b^0c^0$ for both SIO and SRO around the lower interface.

In conclusion, this work brings at least threefold advancements. First, we understood how interfacial DMI could be modified by changes in octahedral tilts in the highly-debated SRO-SIO system, whose crystal structures are intrinsically non-centrosymmetric. Such information of DMI is richer and opens more opportunities than metallic multilayer systems. Furthermore, we demonstrated the stabilization of THE in freestanding flakes via the ELO approach, which shows trends consistent to micromagnetic simulations of magnetic Skyrmions; The relationship between DMI and octahedral tilts is also discovered, i.e., the $a^-/b^-$ tilt is crucial at the top interface of SRO on SIO after relaxing from tilt-suppression. Such observation of THE in freestanding oxide flakes lay the foundation to enable several magnetic imaging techniques and spectroscopies with probe-beam transmission mode for magnetic textures.

**Methods:**

*Sample fabrications*

STO(001) substrates were etched with buffered HF solution and annealed at 950 °C to achieve a single $TiO_2$-termination. All films involved were grown in pulsed laser deposition chambers. To define the Hall bars for structures A, B and C, the substrates were first patterned by photolithography followed by amorphous AlN deposition at room temperature and base vacuum. This eliminates the possibility of introducing oxygen vacancies into the STO substrates by standard Argon ion milling that might create unwanted complications. Hence, we could exclude analyses of impurity/disorder-induced extrinsic AHE mechanisms such as skew scattering or side-jump[45]. All single-crystalline SRO and SIO films were grown at 650 °C, 100 mTorr of oxygen, 2 J/cm$^2$ of laser energy density, and 5 Hz and 1 Hz laser repetition rates for SRO and SIO, respectively. Such parameter should result in a majority monoclinic-phase[25, 46] of SRO (Glazer notation $a^-b^+c^-$) with minimal <0.2 nm surface roughness, as distinct to the tetragonal-phase ($a^0a^0c^-$). For structures $A_f$, $B_f$ and $C_f$, 100 nm-thick SAO films were grown at 830 °C and base vacuum (<5x10$^{-6}$ Torr), followed by 40 nm-thick STO buffer at 830 °C and 5x10$^{-4}$ Torr to ensure layer-by-layer growth.

*Characterizations*

All Ordinary Hall Effect (OHE) background, which is linear with a magnetic field, has been removed from the Hall Effect data presented in the main text and supplementary figures. Rotation of magnetic field is done using a Quantum Design PPMS sample rotator setup. The magnetic moment is measured by a Quantum Design MPMS3 SQUID-VSM. Since the contributions of oxygen vacancy to electrical transport and magnetic moment are difficult to quantify, the 40 nm-thick STO buffer is excluded from the conversion of raw data of resistance into resistivity and magnetic moment into emu/cc. XRD was performed in Singapore Synchrotron Light Source (SSLS) by using the STO(001) as the reference crystal system for conversion from real- to reciprocal-space. Freestanding samples were grown no older than one day before the synchrotron XRD measurement and kept in a vacuum seal before measurement so that SAO is avoided absorbing moisture. We treat any data below 25 counts as noise level or possible charge density waves that are out-of-scope here.

*Calculations*

The Vienna *Ab-initio* Simulation Package (VASP) package was employed in three steps [36]. First, structural relaxations were performed on top of input obtained from half-integer HKL

XRD until the forces became smaller than 0.01 eV/Å for determining the most stable interfacial geometries, resulting in small adjustments from the input. Next, the Kohn-Sham equations were solved, with no SOC, to find out the charge distribution of the system's ground state. Finally, SOC was included, and the self-consistent total energies of the system were determined as a function of the $Ru^{4+}$ magnetic moment orientations (cycloidal wavevector $q$), which were controlled by using the constrained method implemented in VASP.

**Conflicting Interests:**

The authors declare no conflicting interests.

**Acknowledgments:**

This research is supported by the Agency for Science, Technology and Research (A*STAR) under its Advanced Manufacturing and Engineering (AME) Individual Research Grant (IRG) (A1983c0034) and the National Research Foundation (NRF) of Singapore under its NRF-ISF joint program (Grant No. NRF2020-NRF-ISF004-3518).

**Figures:**

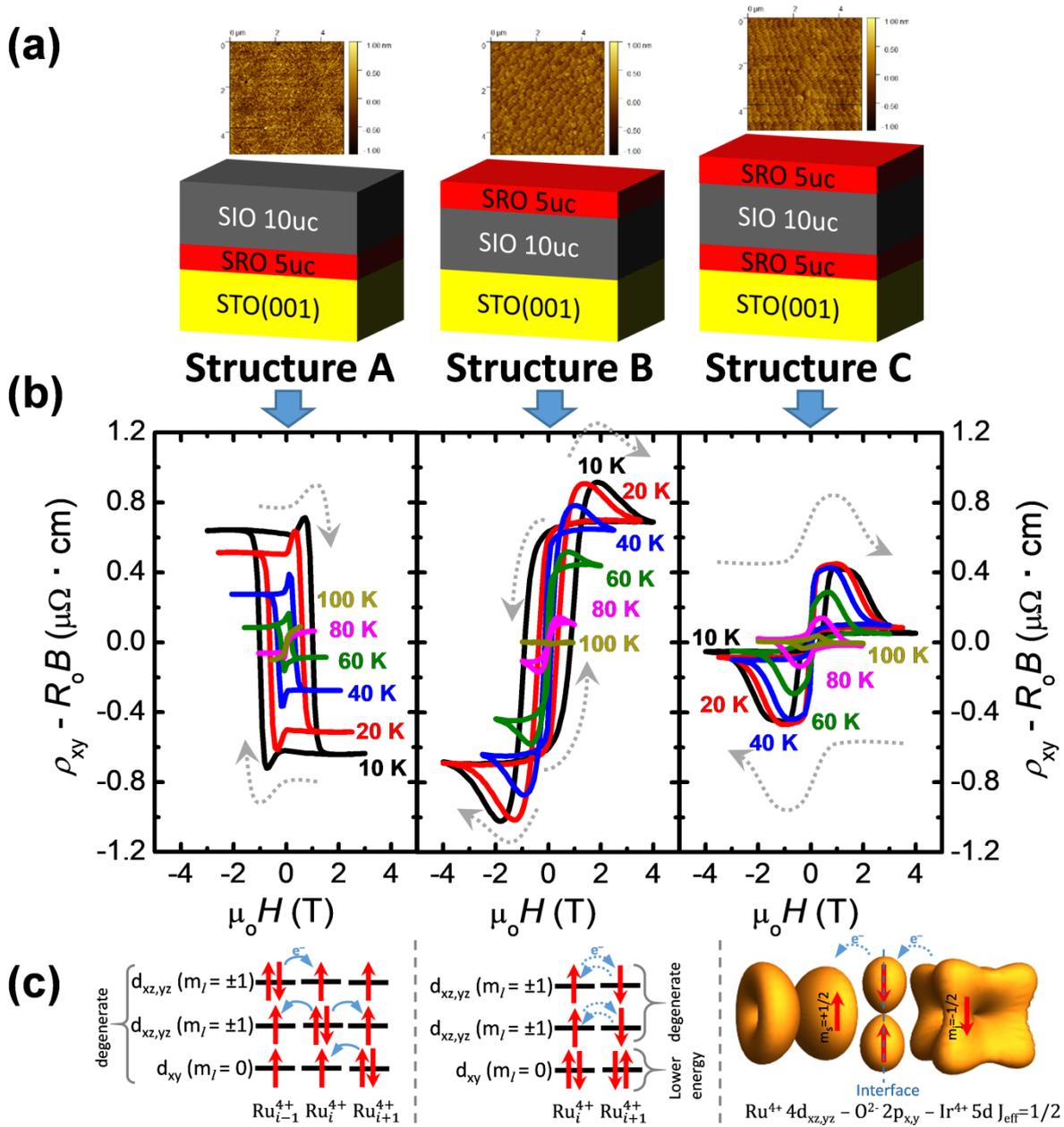

**Figure 1: Hall Effect data of the epitaxial basis. (a)** Schematics of structures A, B and C and their surface topography. **(b)** The corresponding Hall Effect data from the three epitaxial structures, with dotted lines indicating the hysteresis direction. **(c)** Illustrations of (oxygen-mediated) threefold-degenerate ferromagnetic minority-spin double-exchange (left panel) and "Ferro-orbital" antiferromagnetic superexchange (middle panel) between neighboring $Ru^{4+}$ ions in perovskite environment. Right panel: Proposed $Ru^{4+}/Ir^{4+}$ interfacial ferromagnetic superexchange.

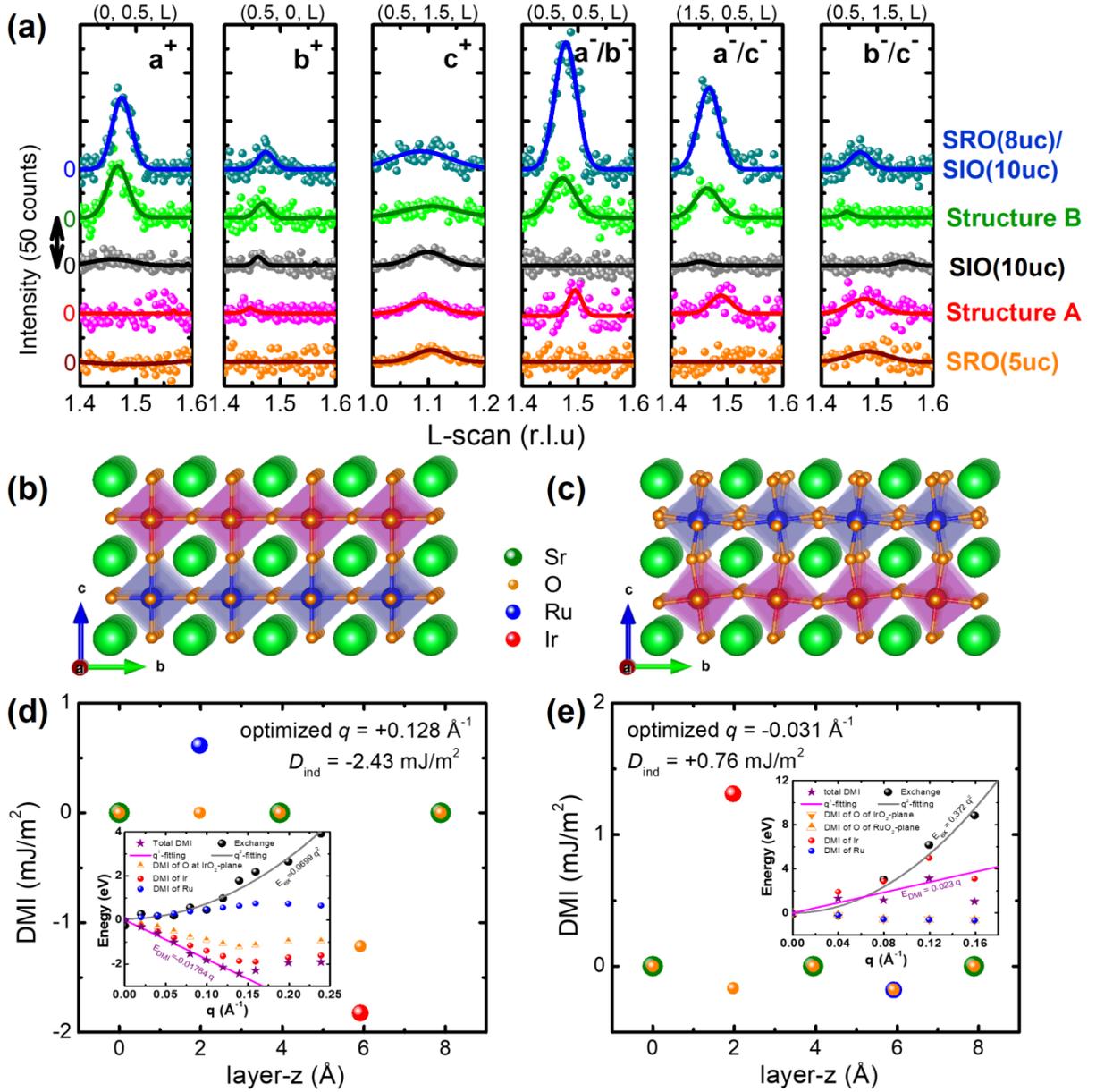

**Figure 2: Experimentally observed structural details for first-principle DMI calculations.** **(a)** Layered-resolved half-integer HKL XRD data for structure A (brown and red curves) and structure B (black and green curves), while solid lines are best-fits. Altered structure B with thicker top SRO is also shown for comparison. **(b) & (c)** Atomic drawings for a two-unit-cell-thick slab representing the interface of structures A and B respectively after relaxation in DFT calculations. Due to octahedral tilts, the simulated volume of (c) is four times larger than (b). **(d) & (e)** The atomic-layer resolved DMI at optimized $q$ for structure A and B, respectively, where the colors on data points follow that of atom labels in (b,c). The $q$-dependent variations of exchange and DMI energies are also shown in insets with fitting coefficients $C_{ex}$ and $D_{ind}$ indicated.

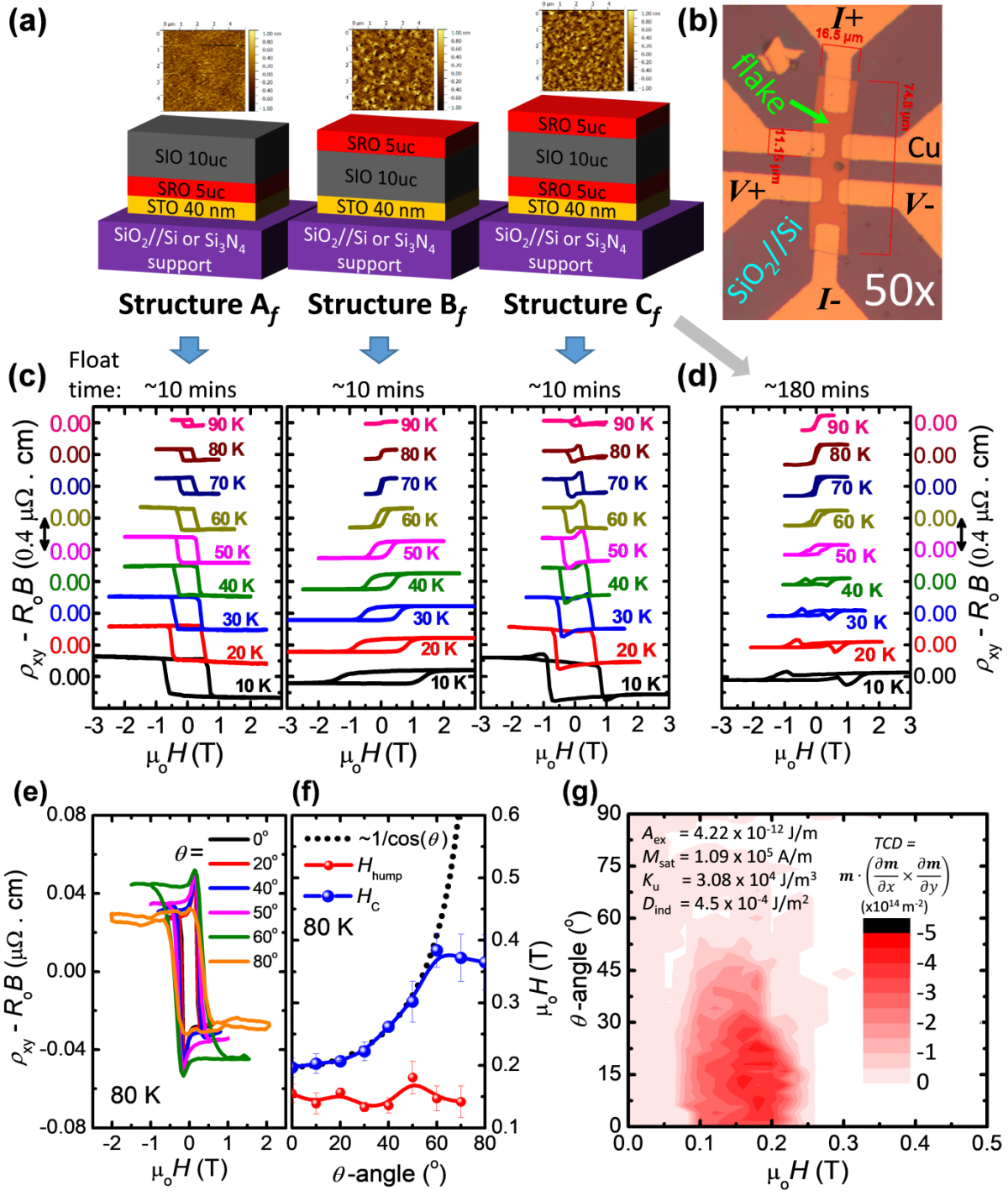

**Figure 3: Hall Effects of the freestanding flakes and micromagnetic simulations. (a)** Schematics of structure $A_f$, $B_f$ and $C_f$ with topography attached. **(b)** Optical microscopy image of a flake patterned with Cu electrodes. Hall Effect loops for **(c)** structure $A_f$ (left), $B_f$ (middle) and $C_f$ (right) after minimal float time, and **(d)** for structure $C_f$ after long float time. The data in (c) and (d) are shifted vertically for clarity. **(e)** Hall Effect loops with varying $\theta$ and their evolutions of $H_{hump}$ and $H_C$ are summarised in **(f)**. **(g)** $\theta$-$\mu_o H$ mapping of $TCD$ by MUMAX$^3$ simulations of Néel-type Skyrmions.

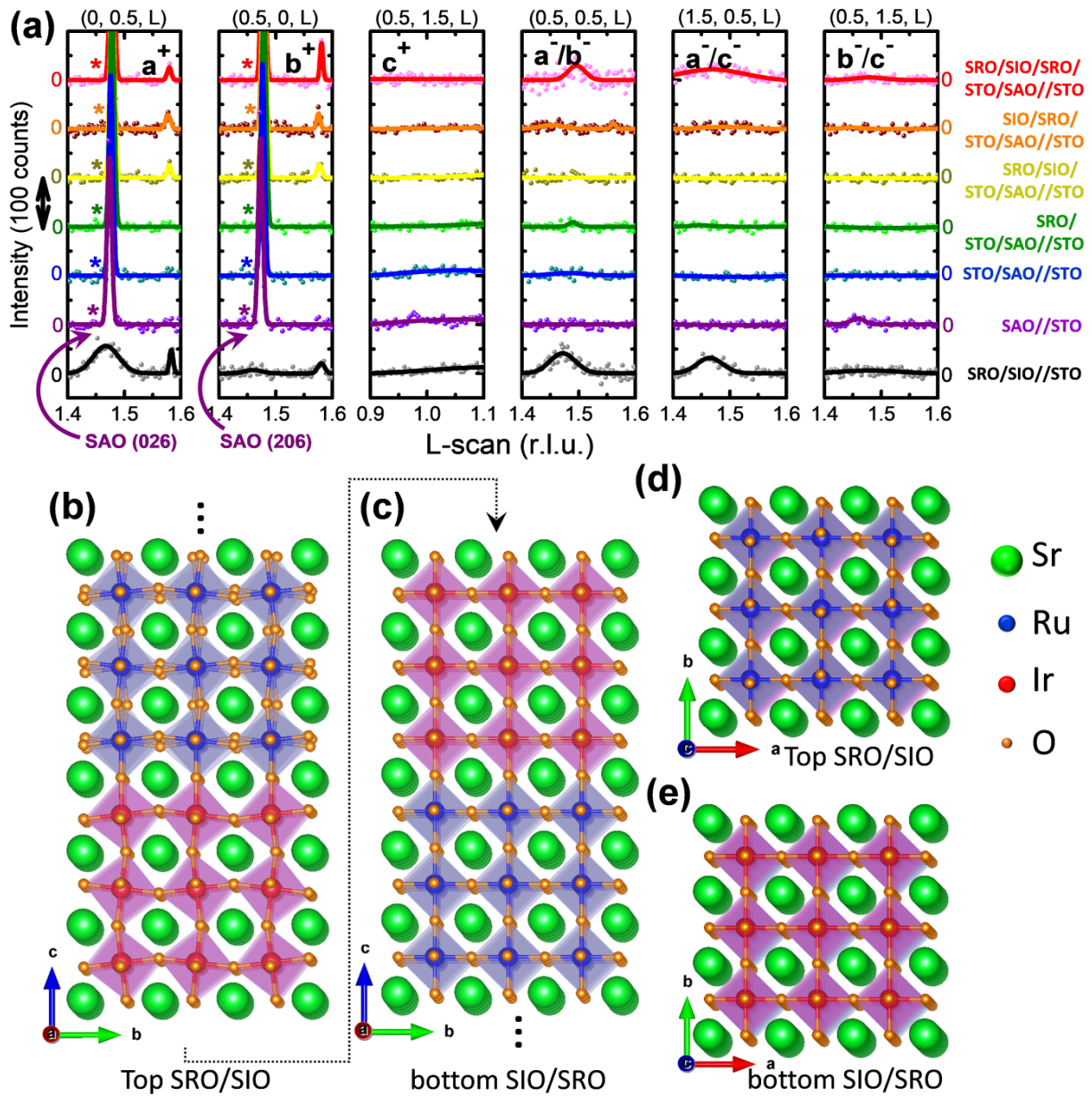

**Figure 4: Half-integer HKL X-ray Bragg diffraction and atomic structure understanding of SAO-buffered heterostructures. (a)** XRD data for investigating $a^+$, $b^+$, $c^+$, $a^-/b^-$, $a^-/c^-$, and $b^-/c^-$ tilt for layer-resolved structure $C_f$ (red to purple curves) without water dissolution in comparison with structure B (black curves), while solid lines are best-fits. Film thicknesses are the same as described in Figures 1 and 3, while 'asterisks' mark the (026) and (206) peaks of SAO. Data are shifted vertically for clarity. **(b)-(e)** Atomic structure of the freestanding SRO/SIO/SRO illustrated by Vesta drawings. The dotted arrow connects (b) and (c), which are side-view along the a-direction, while (d) and (e) are top-view along the c-direction corresponding to (b) and (c), respectively.

Supporting Information

**Oxygen Octahedral Tilt Controlled Topological Hall Effect in Freestanding SrRuO$_3$/SrIrO$_3$ Heterostructures**

Z. S. Lim, A. K. H. Khoo, Z. Zhou, G. J. Omar, P. Yang, R. Laskowski, A. Ariando*

In Figure S1a, for structure A with SRO on STO(001), one RHEED intensity oscillation can be seen followed by a transition into step flow growth mode with no further oscillations, but the surface is smooth enough to support layer-by-layer of the subsequent SIO. Likewise, for structure B, the SIO on STO(001) shows many RHEED intensity oscillations followed by SRO showing no oscillation indicating SRO begins step flow mode immediately. In Figure S1b, both the Curie temperature ($T_C$) and saturation magnetization ($M_{sat}$) of structure B (130 K and ~1.0 μ$_B$/Ru) are higher than that of structure A (100 K, and ~0.52 μ$_B$/Ru), although having the same SRO thickness. Both structures have clear perpendicular magnetic anisotropy. The blocking temperatures ($T_B$) for structure B at both out-of-plane and in-plane fields are 117 K for structure B, while $T_B$ for structure A is higher at the out-of-plane field (54 K) compared to the in-plane field (28 K). The linear resistivity versus temperature ($\rho_{xx}$-$T$) of structures A and B are shown in Figure S1c left panel. Structure B has stronger ferromagnetism but is more insulating because the SIO(10uc) dominates the linear resistivity in both structures. Since SIO in structure B is grown on STO(001) substrate, it has more severe localization. In short, the material grown directly onto STO(001) will suffer from weakened bulk properties – weaker ferromagnetism for SRO (structure A) and higher resistivity for SIO (structure B). In Figure S1c right panel, the modified structure A* having thinner bottom SRO (4uc) recovers the positive Anomalous Hall Effect (AHE), as distinct to structure A in Figure 1b but consistent to reference [23].

Figures S2a-b show that the optical microscopy image of a freestanding flake of structure C$_f$ scooped onto SiO2 support will be curled if the STO buffer's thickness on SAO is 100 nm but flat for STO thickness below 40nm. Figure S3c,d show the C$_f$ flakes exfoliated by dicing tapes with cracks but capable of generating reliable magnetometry data. Figure S3e-f shows that a flake as wide as the whole piece of the substrate with very few cracks was lifted-off by using LSMO as the sacrificial layer in place of SAO and its corresponding KI+HCl+H$_2$O etchant. However, its magnetic signal is completely destroyed and becomes highly insulating.

For main text Figure 3g and supporting Figure S3, the hysteretic behavior of topological charge density, TCD $= \hat{\boldsymbol{m}} \cdot \left(\frac{\partial \hat{\boldsymbol{m}}}{\partial x} \times \frac{\partial \hat{\boldsymbol{m}}}{\partial y}\right)$ is simulated via MUMAX$^3$, using logical magnetic parameters of SRO at 80 K. These parameters are not extracted from the magnetometry of freestanding trilayer structure C$_f$ due to extra signal from the proximity induced magnetism of iridate and oxygen vacancies. Instead, we use $J_{\text{ex}} = \frac{3 k_B T_c}{2 j (j+1)}$ to estimate the exchange constant, where j is the total angular momentum quantum number. 4d$^4$ electronic configuration is arguably lying between the weak SOC (l-s coupling) and strong SOC (j-j coupling) regimes, with S=1, L=1, and **J**=**L**+**S**=1 because the two vectors are neither parallel nor antiparallel. Using bulk values, $M_{\text{sat}}$ ~1.6 μ$_B$/Ru, j=1, $T_c$=160 K, $J_{\text{ex}} = 3k_B T_C/4$, hence $A_{\text{ex}} = J_{\text{ex}}/a = 4.22 \times 10^{-12}$ J/m where a is the pseudocubic lattice parameter 3.94 Å. To account for weakened magnetism in the ultrathin regime, we assume $T_C = 100$ K and $M_{\text{sat}}$=1.0 μ$_B$/Ru at 0 K. Knowing that temperature-dependent magnetization usually takes the form of $M/M_{\text{max}} = \tanh\left(\frac{M/M_{\text{max}}}{T/T_c}\right)$ from the mean-field generalization to the Langevin's paramagnetism, at 80 K, $T/T_c = \frac{M/M_{\text{max}}}{\tanh^{-1}(M/M_{\text{max}})} = 0.8$ can be obtained when $M/M_{\text{max}} = 0.710412$, hence $M_{\text{sat}} = 0.71$ μ$_B$/Ru $= 1.09 \times 10^5$ A/m at 80 K, where $M_{\text{max}}$ refers to $M_{\text{sat}}$ at 0 K. Noting exchange length $l_{\text{ex}} = \sqrt{\frac{2 A_{\text{ex}}}{\mu_o M_{\text{sat}}^2}}$=23.86 nm, hence we use a cell size of 4 nm < $l_{\text{ex}}$. The uniaxial magnetic anisotropy $K_u$ is arbitrarily scaled down to $3.08 \times 10^4$ J/m$^3$. Gilbert damping constant is set to 0.1, while the backward/forward field sweep calculations are always initialized with m=Uniform(0,0,±1) so that the emergence of Skyrmions and *TCD* are always hysteretic. Combining with DMI $D_{\text{ind}} = 4.5 \times 10^{-4}$ J/m$^2$ induced at the SRO/SIO interface, we obtain a *TCD* variation within a field range that matches well with the observed Hall Effect hump features, i.e., it reaches the peak at μ$_o$H~±0.14 T and vanishes at μ$_o$H~±0.3 T. Subsequently the same parameters are used to generate the θ-dependent trend as shown in the main text Figure 3g, at μ$_o$H$_{\text{peak}}$ = 0.14 T, corresponding to Néel-type Skyrmions' diameter of ~100 nm.

In Figure S4a,b, we compare two cases of Néel-Skyrmions with hysteretic and non-hysteretic *TCD* by MUMAX$^3$ simulations. This part of the work shows the fallibility of the argument that minor loops in Hall Effect (being reversible or non-reversible) can be used to distinguish THE from real magnetic Skyrmions from the bi-AHE phenomenon produced by the coexistence of two types of regular domains with opposite-sign AHE loops and *k*-space Berry curvatures. In both cases, the backward/forward magnetic field sweep is done by always using m=Uniform(0,0,±1), representing domain evolution after saturation. In Figure S4a, using an exemplary set of parameters derived from magnetometry of structure B (Figure S1b): $A_{ex}$=2.63x10$^{12}$ J/m, $M_{sat}$=198.5 kA/m, $K_U$=24.8 kJ/m$^3$ (from $M$-$H$ loops) and an arbitrary interfacial DMI $D_{ind}$=0.4 mJ/m$^2$, hysteretic $TCD$ appears at the small negative field after positive large field saturation and vice versa, i.e., $H_{peak}$=∓0.1 T after ±1T saturation, but is absent at the ±0.1 T after ±1T saturation. Then, we simulate a minor loop field sweep by using the solved magnetic state at $H_{peak}$ as the initialization for the subsequent field sweep in the reversed direction towards zero field. The resulting $TCD$ and magnetic state at zero field become the same as $H_{peak}$, implying the minor loop field sweep field result is "non-reversible". However, in Figure S4b, with the same parameter $A_{ex}$, $M_{sat}$ and $K_U$ but higher $D_{ind}$=1.85 mJ/m$^2$, non-hysteretic $TCD$ appears at small fields of both polarity ($H_{peak}$=±1 T) after large field (±4 T) saturation regardless of polarity. Such non-hysteretic THE is similar to those that appeared in reference [40]. With the same minor loop simulation method as (a), i.e., using the magnetic states at ±$H_{peak}$ as initialization and sweeping the field back to zero, we see that the $TCD$ drops to near zero, corresponding to the cycloidal stripe phase. This implies that the reversibility of minor loops of Hall measurement is not a convincing tool/judgment to distinguish magnetic Skyrmions from regular domains without magnetic imaging since both simulations involve topological Skyrmions. In Figure S4c, after application of a strong in-plane field $H_y$ on top of a particular out-of-plane field $H_z$* supporting Skyrmion-lattice, only the $q_y$ Bloch-type helical and $q_x$ Néel-type cycloidal magnons would survive.

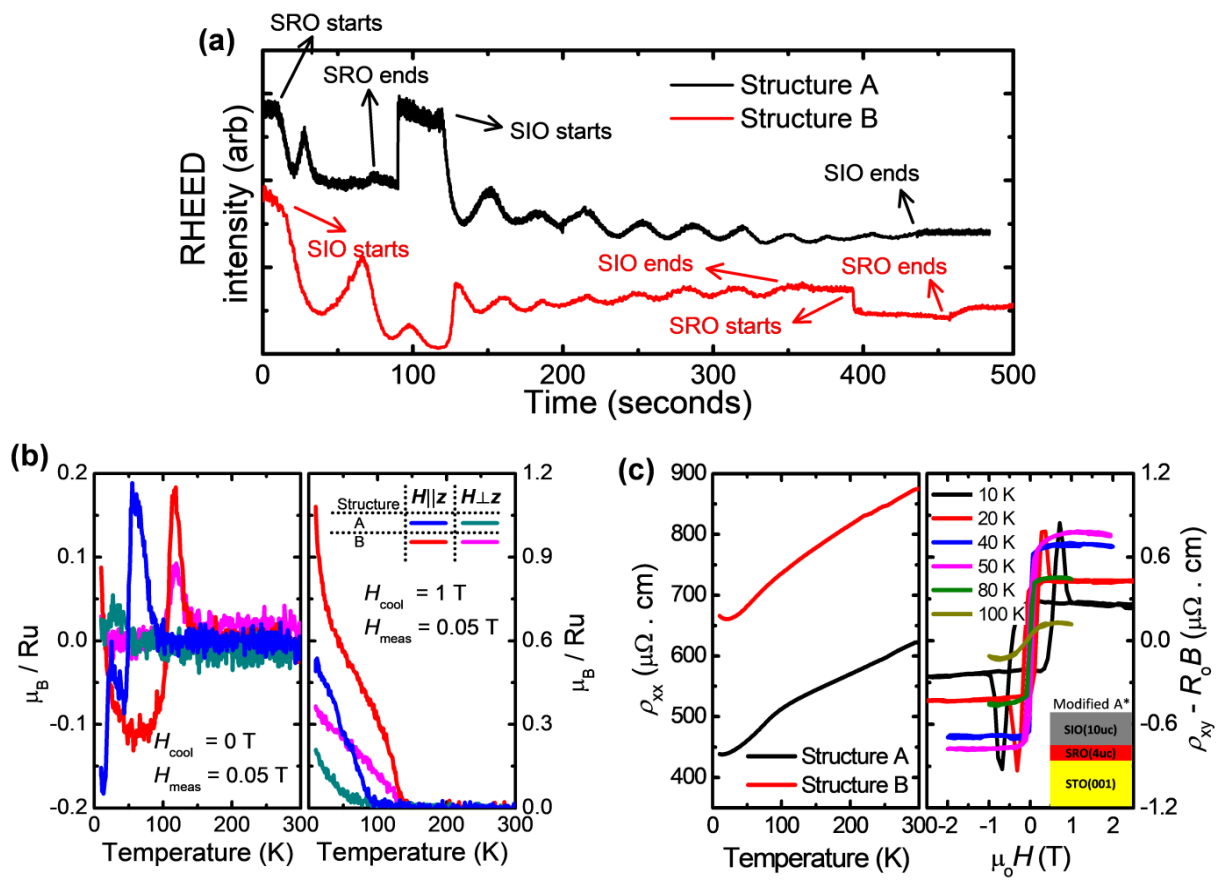

**Supporting Figure S1:** (a) RHEED intensity oscillations during film growth of structures A and B. (b) *M-T* curves for zero-field cooling (left) and 1 T-field cooling (right) for structures A and B, measured by 0.05 T field during warming. (c) Left: $\rho_{xx}$-*T* of structures A and B. Right: Hall Effect from the modified structure A* by reducing the bottom SRO thickness to 4uc.

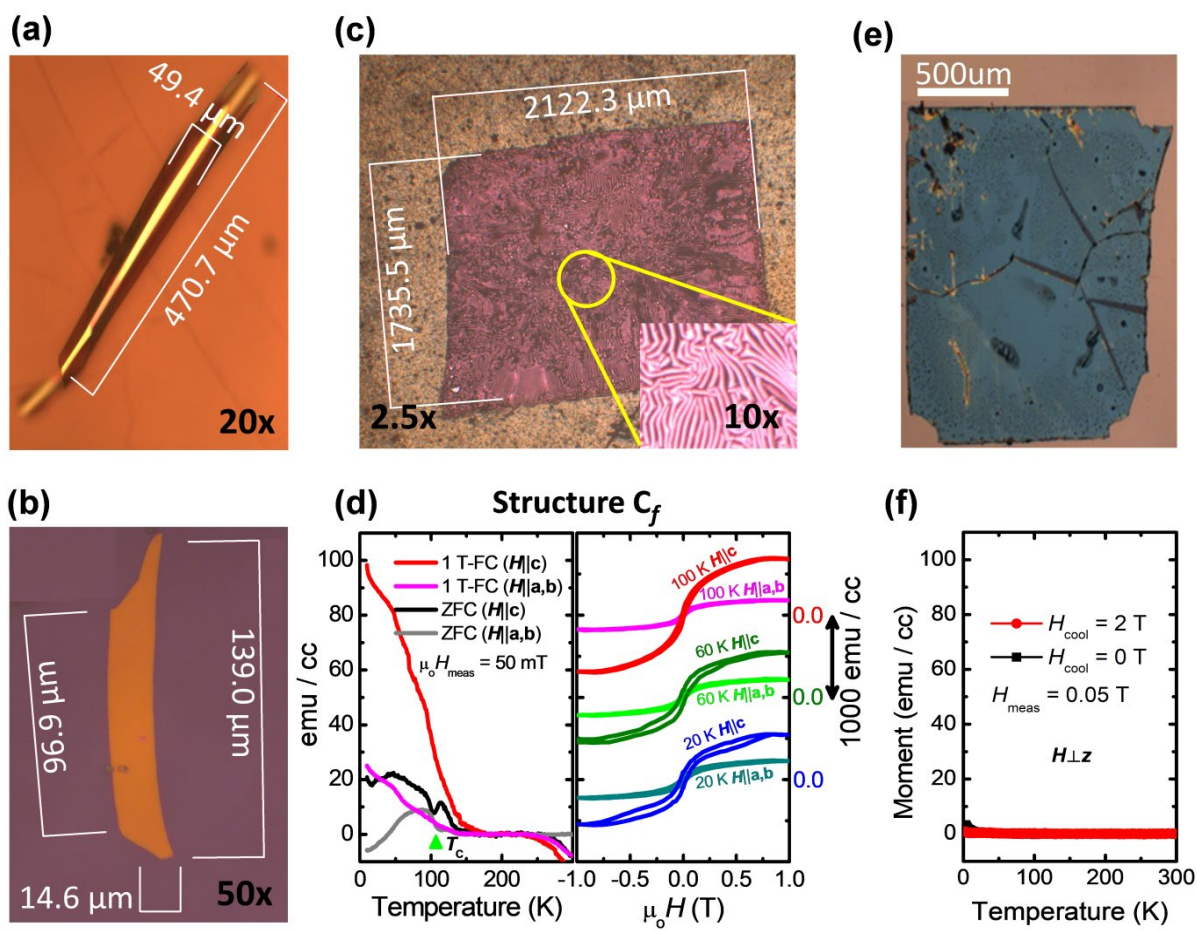

**Supporting Figure S2: Flakes' morphology and magnetic properties.** Optical microscopy images of curled-up roll **(a)** and flat flake **(b)**, adhered on $SiO_2$//Si support. **(c)** Optical microscopy image of an aggregate of cracked flakes exfoliated by a polyimide (Kapton) tape, to be used for generating the *M-T* (left) and *M-H* curves (right) shown in **(d)**. The *M-H* curves of different temperatures are shifted vertically for clarity. **(e)** A large continuous flake of structure $C_f$ with LSMO sacrificial layer floated by an $HCl+KI+H_2O$ solution and transferred onto $SiO_2$, without any tape support, which is found to be devoid of any ferromagnetic signal in magnetometry **(f)**.

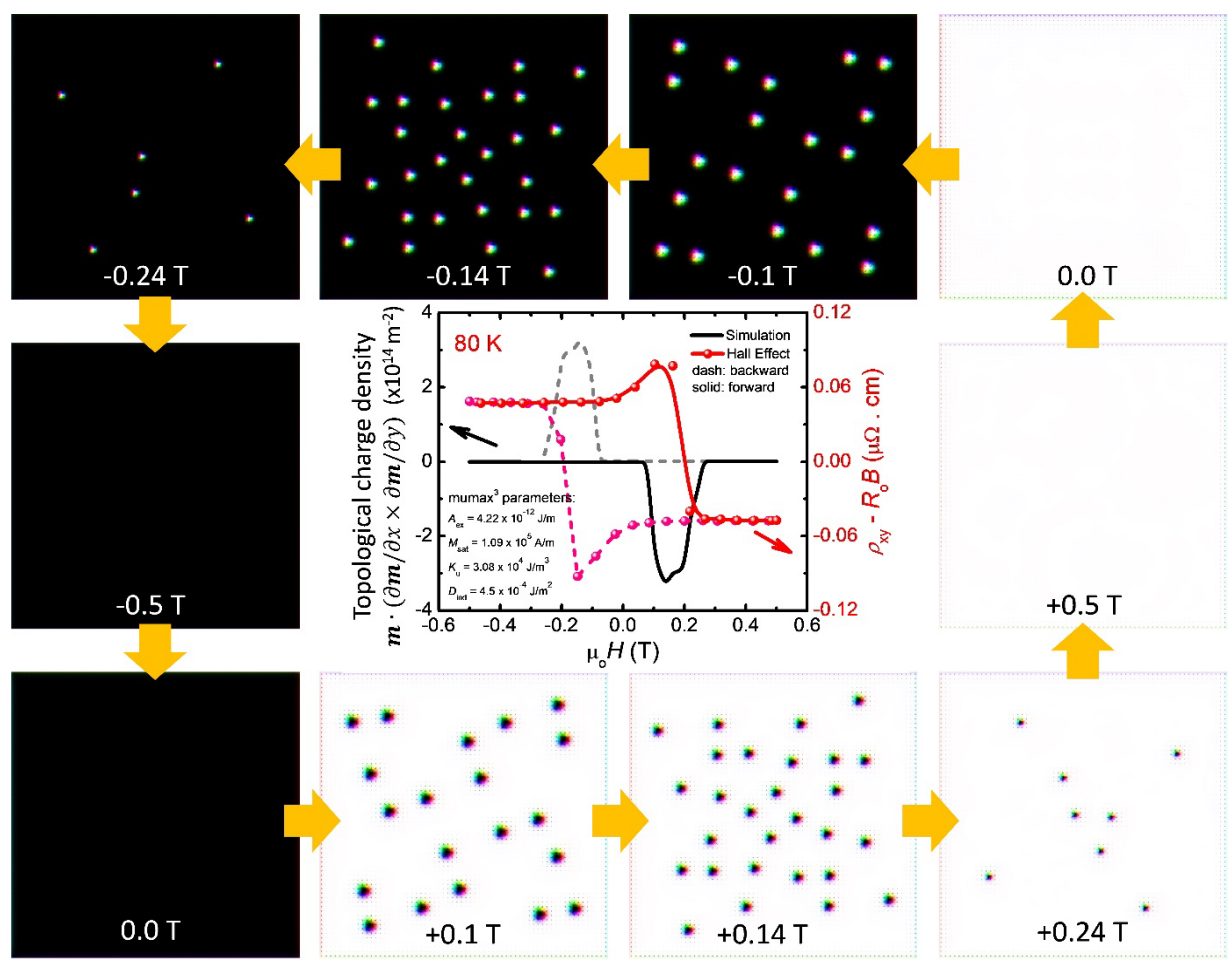

**Supporting Figure S3:** MUMAX³-simulated field-sweep result of the *TCD* showing hysteretic behavior overlapped with the experimental Hall Effect for comparison. The surrounding snapshots show the corresponding magnetic textures at various magnetic fields.

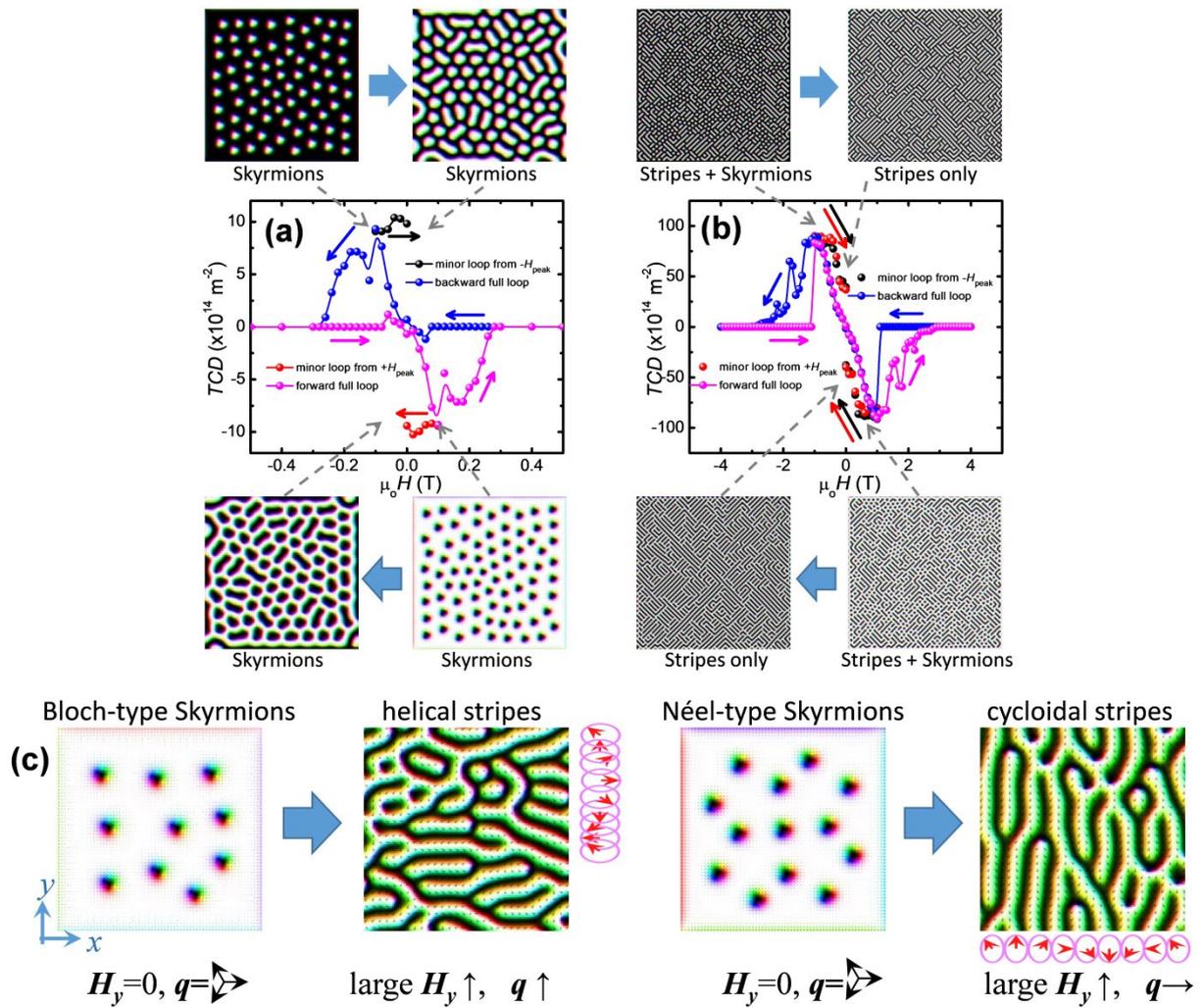

**Supporting Figure S4:** MUMAX$^3$-simulated full and minor *TCD* loops for **(a)** hysteretic (lower DMI) and **(b)** non-hysteretic (higher DMI) THE, respectively. The magnetic states corresponding to the $H_{peak,TCD}$ and zero field are also shown in the insets. (c). Simulated evolution of magnetic states in thin films before and after applying a strong in-plane field $H_y$ for both Bloch- and Néel-Skyrmions. The arrows indicate the directions of magnon wavevectors.